\documentclass[twocolumn,preprintnumbers,showkeys,amsmath,amssymb]{revtex4}
\usepackage{graphicx}
\usepackage{amssymb}

\usepackage{color}

\begin{document}

\title{An optical implementation of quantum bit commitment using
infinite-dimensional systems}
\author{Guang Ping He}
\email{hegp@mail.sysu.edu.cn}
\affiliation{School of Physics, Sun Yat-sen University, Guangzhou 510275, China}

\begin{abstract}
Unconditionally secure quantum bit commitment (QBC) was widely believed to
be impossible for more than two decades. But recently, based on an anomalous
behavior found in quantum steering, we proposed a QBC protocol which can be
unconditionally secure in principle. The protocol requires the use of
infinite-dimensional systems, therefore it may seem less feasible in
practice. Here we propose a quantum optical method based on Mach-Zehnder
interferometer, which gives a very good approximation to such
infinite-dimensional systems. Thus, it enables a proof-of-principle
experimental implementation of our protocol, which can also serve as a
practically secure QBC scheme.
\textcolor{black}{Other multi-party cryptographic protocols such as quantum coin tossing can be built upon it too. Our approach also reveals a relationship between infinity and non-locality, which may have an impact on the research of fundamental theories.}
\end{abstract}

\keywords{secure multi-party computation; bit commitment; coin
tossing; Mach-Zehnder interferometer; quantum steering.}

\maketitle


\section{Introduction}

Quantum cryptography has achieved great success in many fields such as key
distribution \cite{qi365}, but there are still other cryptographic problems
remain unconquered. Bit commitment (BC) \cite{qi43} is known to be an
essential building block for coin tossing \cite{qi365}, oblivious transfer\
\cite{qbc9,qbc185}, and even more complicated multi-party secure computation
protocols \cite{qi139}. Unfortunately, since 1996, 
\textcolor{black}{people started to realize that unconditionally secure quantum BC (QBC) is hard to achieve. The cheating strategy against the QBC protocol in Ref. \cite{qi43} was first proposed in
Ref. \cite{qi74}. Shortly after, it was further asserted that all QBC protocols are not unconditionally secure in principle \cite{qi24,qi23}. Later, Refs. \cite{qi56,qi105,qi58,qbc28,qi611,qbc121} reviewed the
original no-go proof, with some examples of
insecure protocols given in Refs. \cite{qi56,qi611}. Ref. \cite{qi58} also
extended the proof to cover ideal quantum coin tossing (QCT). More
examples on how to break some promising BC protocols at that time
were provided too \cite{qi82,qi47}. Refs. \cite{qbc183,qi581,qbc158,qbc25} proved the impossibility of some types of BC with slightly different security criterion. Refs. \cite%
{qi147,qbc49,qi101,qbc3,qbc31} gave quantitative studies on the security bounds of QBC, with Ref. \cite{qbc49} focused on the protocol in Ref. \cite{qi365} while
Refs. \cite{qi101,qbc3} focused on another class of protocols. A
very lengthy proof was first presented in the Heisenberg picture \cite{qi323},
then shortened and rephrased in the Schr\"{o}dinger picture \cite{qi715}. The validity of the no-go result was also studied in a world subject to
superselection rules \cite{qbc36,qi610,qbc35} or an epistemic local hidden variable theory \cite{qbc146}, as well as for QBC associated with
secret parameters \cite{qi240,qi283} or secret probability distributions \cite{qbc208}, or when the participants are
restricted to use Gaussian states and operations only \cite{qi714}.
Refs. \cite{qbc140,qbc255} attempted to deduce the impossibility of QBC from the no-masking theorem. Ref. \cite{qbc156} studied the security of BC under the relativistic setting.
Other efforts include Refs. \cite{qbc12,qbc32,qbc148,qbc162,qbc182,qbc227}, which tried to proved the no-go theorem with alternative approaches.
These results},
known as the Mayers-Lo-Chau (MLC) no-go theorem, \textcolor{black}{were} widely
accepted despite of some attempts towards secure QBC (e.g.,
\textcolor{black}{the references in Refs. \cite%
{HePRA,HeJPA,HeQIP,HeBook,qbc75,qbc157,qbc273}}), and \textcolor{black}{were}
considered as putting a serious drawback on the potential of quantum
cryptography.

Nevertheless, the cheating strategy in all these no-go proofs is based on
the Hughston-Jozsa-Wootters (HJW) theorem \cite{qi73}, a.k.a. the Uhlmann
theorem \cite{qbc8,qbc48}. Recently, it was found that in
infinite-dimensional systems, there exists a specific form of quantum states
to which the HJW theorem does not apply \cite{HeSteering}. Based on this
finding, we proposed a QBC protocol and proved theoretically that it remains
secure against the cheating strategy in the no-go proofs \cite{HePRSA}.
Therefore, implementing the protocol in practice will be of great
significance as it can re-open the venue to many useful multi-party secure
computation protocols that was once closed by the MLC no-go theorem.

As pointed out in Ref. \cite{HePRSA}, since the protocol requires
infinite-dimensional systems, the implementation may be very hard if we want
to use physical systems with an infinite number of energy levels, because it
may imply an infinitely high energy. To circumvent the problem, here we use
the arrival time of photons as a trick, so that the infinite-dimensional
systems can be realized using simple optical devices. Consequently, the QBC
protocol in Ref. \cite{HePRSA} can be implemented with Mach-Zehnder (MZ)
interferometer, which is within the capability of currently available
technology.

\section{Results}

\subsection{The theoretical description of the protocol}

Let us begin with a brief review on the definition of BC and the theoretical
scheme in Ref. \cite{HePRSA}. BC is a two-party cryptography between Alice
and Bob, which includes the following phases. In the commit phase, Alice
decides the value of the bit $b$ that she wants to commit, and sends Bob a
piece of evidence, e.g., some quantum states. Later, in the unveil phase,
Alice announces the value of $b$, and Bob checks it with the evidence. The
interval between the commit and unveil phases can be called the holding
phase. An unconditionally secure BC protocol needs to be both binding (i.e.,
Alice cannot change the value of $b$ after the commit phase) and concealing
(Bob cannot know $b$ before the unveil phase).

Since whether QBC can be unconditionally secure in principle is a very
important theoretical problem, here we only consider the ideal case without
transmission errors, detection loss, dark counts, or other practical
imperfections. In Ref. \cite{HePRSA}, the following protocol was proposed.

\bigskip

\textit{Our theoretical QBC protocol:}

\textit{The commit phase:}

\textit{(i) Alice decides on the value of }$b$\textit{\ (}$b=0$\textit{\ or }%
$1$\textit{)\ that she wants to commit. Then for }$j=1$\textit{\ to }$s$%
\textit{:}

\textit{She randomly picks an integer }$i_{j}\in \{1,2,...,\infty \}$\textit{%
, and sends Bob a quantum register }$\Psi _{j}$\textit{, which\ is an
infinite-dimensional system prepared in the state }$\psi
_{i_{j}}^{b}=(\left\vert 0\right\rangle +(-1)^{b}\left\vert
i_{j}\right\rangle )/\sqrt{2}$\textit{.}

\textit{That is, if }$b=0$\textit{\ she randomly picks a state from the set }%
\begin{equation}
\left\{ \psi _{i}^{0}\equiv \left\vert \phi _{i+}\right\rangle =\frac{1}{%
\sqrt{2}}(\left\vert 0\right\rangle +\left\vert i\right\rangle
),i=1,...,n-1\right\} ,  \label{state+}
\end{equation}%
\textit{or if }$b=1$\textit{\ she randomly picks a state from the set}%
\begin{equation}
\left\{ \psi _{i}^{1}\equiv \left\vert \phi _{i-}\right\rangle =\frac{1}{%
\sqrt{2}}(\left\vert 0\right\rangle -\left\vert i\right\rangle
),i=1,...,n-1\right\} ,  \label{state-}
\end{equation}%
\textit{where }$\left\vert 0\right\rangle $\textit{, }$\left\vert
1\right\rangle $\textit{, }$\left\vert 2\right\rangle $\textit{, ... , }$%
\left\vert i\right\rangle $\textit{, ... are orthogonal to each other,\ and }%
$n\rightarrow \infty $\textit{.}

\textit{Note that in each round, }$i_{j}$\textit{\ is independently chosen,
while }$b$\textit{\ remains the same for all }$j$\textit{.}

\textit{The holding phase:}

\textit{(ii) Bob stores these }$s$\textit{\ quantum registers unmeasured.}

\textit{The unveil phase:}

\textit{(iii) Alice announces the values of }$b$\textit{\ and all }$i_{j}$%
\textit{\ (}$j=1,...,s$\textit{).}

\textit{(iv) Bob tries to project each }$\Psi _{j}$\textit{\ into the state }%
$\psi _{i_{j}}^{b}=(\left\vert 0\right\rangle +(-1)^{b}\left\vert
i_{j}\right\rangle )/\sqrt{2}$\textit{. If the projections are successful
for all registers, Bob accepts Alice's commitment. Else if any of the
projections fails, Bob concludes that Alice cheated.}

\bigskip

The key reason that this protocol can be unconditionally secure, is the
specific forms of the states in Eqs. (\ref{state+}) and (\ref{state-}). In
general, the cheating strategy in the no-go proofs
\textcolor{black}{\cite%
{qi74,qi24,qi23,qi56,qi105,qi58,qbc28,qi611,qbc121,qi82,qi47,qbc183,qi581,qbc158,qbc25,qi147,qbc49,qi101,qbc3,qbc31,qi323,qi715,qbc36,qi610,qbc35,qbc146,qi240,qi283,qbc208,qi714,qbc140,qbc255,qbc156,qbc12,qbc32,qbc148,qbc162,qbc182,qbc227}}
can be successful in most QBC protocols using other forms of quantum states for the
following reason. Suppose that honest Alice is supposed to send Bob the
state $\psi ^{\prime 0}$\ ($\psi ^{\prime 1}$) if she wants to commit $b=0$\
($b=1$), where $\psi ^{\prime 0}$\ ($\psi ^{\prime 1}$) is picked from a set
of states described by the density matrix $\rho _{0}^{\beta }$ ($\rho
_{1}^{\beta }$). Since an unconditionally secure QBC protocol needs to be
concealing against dishonest Bob, there should be%
\begin{equation}
\rho _{0}^{\beta }\simeq \rho _{1}^{\beta }  \label{concealing}
\end{equation}%
so that Bob cannot discriminate the state himself. Then the HJW theorem
applies. That is, dishonest Alice can begin the QBC protocol by preparing
the system $\alpha \otimes \beta $\ in such a state that $\beta $\ alone has
density matrix $\rho _{0}^{\beta }$. Then she skips the measurement in the
commit phase so that $\alpha $ and $\beta $\ remain entangled. In the unveil
phase, since Eq. (\ref{concealing}) is satisfied, according to the HJW
theorem there exist two measurements $M_{0}$ and $M_{1}$\ on $\alpha $, such
that if Alice applies $M_{0}$ ($M_{1}$) on $\alpha $, then $\beta $\ will
collapse to a state belonging to the set described by $\rho _{0}^{\beta }$ ($%
\rho _{1}^{\beta }$). Therefore, Alice can unveil $b$ as whatever value she
likes in the unveil phase by choosing between the two measurements $M_{0}$
and $M_{1}$.

However, in our protocol the two sets of states take the forms in Eqs. (\ref%
{state+}) and (\ref{state-}). Suppose that dishonest Alice prepares a
bipartite system $\alpha \otimes \beta $ in the state%
\begin{equation}
\left\vert \Omega \right\rangle =\frac{1}{\sqrt{n-1}}\sum_{i=1}^{n-1}\left%
\vert \alpha _{i+}\right\rangle \left\vert \phi _{i+}\right\rangle
\label{omega}
\end{equation}%
so that she can cheat using the strategy in the no-go proofs. Here $%
\{\left\vert \alpha _{i+}\right\rangle ,i=0,...,n-1\}$ is an orthonormal
basis of the $n$-dimensional system $\alpha $. Let $\rho _{0}^{\beta }$ and $%
\rho _{1}^{\beta }$ be the density matrices corresponding to the sets $%
\left\{ \psi _{i}^{0}=\left\vert \phi _{i+}\right\rangle \right\} $\ and
\textcolor{black}{$%
\left\{ \psi _{i}^{1}=\left\vert \phi _{i-}\right\rangle \right\} $},
respectively. As shown in Ref. \cite{HeSteering}, when $n\rightarrow \infty $,
\color{black} Eq. (\ref{concealing}) is satisfied,
so that it seems to meet the requirement of the HJW theorem.
Now let us see what happens if Alice wants to cheat.

Surely, if she wants to unveil $b=0$, all she needs is simply to use $\{\left\vert
\alpha _{1+}\right\rangle ,\left\vert \alpha _{2+}\right\rangle
,..., \\
\left\vert \alpha _{(n-1)+}\right\rangle \}$ as the basis of the measurement
$M_{0}$, and applies it on her system $\alpha $, which will make Bob's system $\beta $ collapse into one of the state in $\left\{ \left\vert \phi _{i+}\right\rangle \right\} $\ so that she can complete the protocol without being caught.
Now the question is whether she can unveil $b=1$ successfully. According to the HJW theorem, there should exist another measurement $M_{1}$ with the basis $\{\left\vert
\alpha _{1-}\right\rangle ,\left\vert \alpha _{2-}\right\rangle
,...,\left\vert \alpha _{(n-1)-}\right\rangle \}$, such that Eq. (\ref{omega}) can be expressed as
\begin{equation}
\left\vert \Omega \right\rangle \simeq \frac{1}{\sqrt{n-1}}\sum_{i=1}^{n-1}\left%
\vert \alpha _{i-}\right\rangle \left\vert \phi _{i-}\right\rangle
\label{omega'}
\end{equation}%
so that Alice's measuring $\alpha $ in this basis will make $\beta $ collapse into a state in $\left\{ \left\vert \phi _{i-}\right\rangle \right\}$.
To find the form of $\left\{ \left\vert \alpha _{i-}\right\rangle \right\}$, following Ref. \cite{HePRSA},
we expand each $\left\vert \phi _{i+}\right\rangle $ in Eq. (\ref{omega}) using $\left\{ \left\vert \phi _{i-}\right\rangle \right\}$, and the result is
\begin{eqnarray}
\left\vert \Omega \right\rangle  &=&\frac{1}{\sqrt{n-1}}\sum_{i=1}^{n-1}%
\sqrt{1-\frac{4}{n^{2}}}\left\vert \tilde{\alpha}_{i-}\right\rangle
\left\vert \phi _{i-}\right\rangle   \nonumber \\
&&+\sqrt{\frac{2}{n}}\left\vert \tilde{\alpha}_{n-}\right\rangle \left\vert
\phi _{n-}\right\rangle ,  \label{omega-}
\end{eqnarray}
where
\begin{equation}
\left\vert \phi _{n-}\right\rangle \equiv \frac{1}{\sqrt{n}}\left(
\left\vert 0\right\rangle +\sum\nolimits_{i=1}^{n-1}\left\vert
i\right\rangle \right) ,  \label{fai n}
\end{equation}%
\begin{equation}
\left\vert \tilde{\alpha}_{n-}\right\rangle \equiv \frac{1}{\sqrt{n-1}}%
\sum_{i=1}^{n-1}\left\vert \alpha _{i+}\right\rangle ,  \label{alfa n-}
\end{equation}%
and%
\begin{equation}
\left\vert \tilde{\alpha}_{i-}\right\rangle \equiv \frac{1}{\sqrt{1-\frac{4}{%
n^{2}}}}\left( \frac{2-n}{n}\left\vert \alpha _{i+}\right\rangle +\frac{2}{n}%
\sum_{i^{\prime }=1,i^{\prime }\neq i}^{n-1}\left\vert \alpha _{i^{\prime
}+}\right\rangle \right)   \label{alfa-}
\end{equation}%
for\ $i=1,...,n-1$.

For a given $i\in \{1,...,n-1\}$, if Alice can project system $\alpha $ to $%
\left\vert \tilde{\alpha}_{i-}\right\rangle $, then Eq. (\ref{omega-}) shows
that system $\beta $ will collapse to%
\begin{eqnarray}
\left\vert \tilde{\phi}_{i-}\right\rangle  &\equiv &c^{\prime }\left[ \frac{%
\sqrt{1-\frac{4}{n^{2}}}}{\sqrt{n-1}}\left( \left\vert \phi
_{i-}\right\rangle +\sum_{i^{\prime }=1,i^{\prime }\neq i}^{n-1}\left\langle
\tilde{\alpha}_{i-}\right. \left\vert \tilde{\alpha}_{i^{\prime
}-}\right\rangle \left\vert \phi _{i^{\prime }-}\right\rangle \right)
\right.   \nonumber \\
&&\left. +\sqrt{\frac{2}{n}}\left\langle \tilde{\alpha}_{i-}\right.
\left\vert \tilde{\alpha}_{n-}\right\rangle \left\vert \phi
_{n-}\right\rangle \right]   \nonumber \\
&=&c^{\prime }\left[ \sqrt{\frac{n^{2}-4}{n^{2}(n-1)}}\left( \left\vert \phi
_{i-}\right\rangle -\frac{4\sum_{i^{\prime }=1,i^{\prime }\neq
i}^{n-1}\left\vert \phi _{i^{\prime }-}\right\rangle }{n^{2}-4}\right)
\right.   \nonumber \\
&&\left. +\sqrt{\frac{2(n-2)}{n(n-1)(n+2)}}\left\vert \phi
_{n-}\right\rangle \right] ,  \label{beta final2}
\end{eqnarray}%
where%
\begin{equation}
c^{\prime }=\sqrt{\frac{n(n-1)(n+2)}{(n^{2}+2)}}.
\end{equation}%
Multiplying $\left\langle \phi _{i-}\right\vert $\ by Eq. (\ref{beta final2}%
), we have
\begin{equation}
\left\langle \phi _{i-}\right. \left\vert \tilde{\phi}_{i-}\right\rangle =%
\sqrt{1-\frac{2n+2}{n^{2}+2}}.  \label{new eq25}
\end{equation}%
In the limit $n\rightarrow \infty $,
$\left\vert \tilde{\phi}_{i-}\right\rangle $ can be arbitrarily close to $\left\vert \phi _{i-}\right\rangle $. Thus, we know that $%
\left\vert \tilde{\alpha}_{i-}\right\rangle $ is the form of $\left\{ \left\vert \alpha _{i-}\right\rangle \right\}$ that we are looking for.

Nevertheless, by taking the limit $n\rightarrow \infty $ in Eq. (\ref{alfa-}), we find
\begin{equation}
\left\vert \tilde{\alpha}_{i-}\right\rangle =-\left\vert \alpha
_{i+}\right\rangle   \label{alfa- infinite} .
\end{equation}%
Consequently, if Alice wants to
collapse $\beta $ into a state in $\left\{ \left\vert \phi _{i-}\right\rangle \right\}$ so that she can unveil $b=1$ successfully, then the
corresponding measurement $M_{1}$ is to measure $\alpha $ in the basis $%
\{-\left\vert \alpha _{1+}\right\rangle ,-\left\vert \alpha
_{2+}\right\rangle ,...,-\left\vert \alpha _{(n-1)+}\right\rangle \}$.
\color{black}
Since the global negative sign before
the state vectors has no physical meaning, the bases of the \textquotedblleft
two\textquotedblright\ measurements $M_{0}$ and $M_{1}$ are actually the
same. Consequently, Alice no longer has the freedom to choose between two
different measurements to alter the value of her committed bit $b$. Thus the
cheating strategy in the no-go proofs fails in our protocol. Please see Ref.
\cite{HePRSA} for the complete security proof.

\subsection{The experimental implementation}

\textcolor{black}{Ref. \cite{HePRSA} was devoted to the problem of whether unconditionally secure QBC is allowed in principle. Thus, it only provided a theoretical description of the protocol without considering the implementation.}
To realize
\textcolor{black}{the} protocol, the most important point is to find a feasible
implementation of the \textcolor{black}{infinite-dimensional} systems. Here we propose a trick to implement
the
\textcolor{black}{infinite-dimensional} system in each round of the protocol using a single photon
only. The experimental apparatus is illustrated in Fig. 1. In each of the $s$
rounds of step (i) of the protocol, Alice sends a single photon either from
the source $S_{0}$ (for sending $\psi _{i}^{0}$) or $S_{1}$ (for sending $%
\psi _{i}^{1}$), \textcolor{black}{then splits it} into two wave packets $\left\vert
x\right\rangle $ and $\left\vert y\right\rangle $ by
\textcolor{black}{the 50:50 non-polarizing beam splitter} $BS_{A}$. $\left\vert x\right\rangle $ is sent directly to Bob via path $X$
while $\left\vert y\right\rangle $ is delayed by the storage ring $SR_{A}$
(which introduces a delay time $\tau $ chosen by Alice secretly) before
sending via path $Y$. At Bob's site, $\left\vert x\right\rangle $ is delayed
by the storage rings $SR_{x}$ and $SR_{B}$. $\left\vert y\right\rangle $ is
delayed by the storage ring $SR_{y}$ which is identical to $SR_{x}$ so that
they introduce the same
\textcolor{black}{amount of} delay time, then meets $\left\vert x\right\rangle $
at the
\textcolor{black}{50:50} beam splitter $BS_{B}$ and interferes. We can see that when the delay
times caused by $SR_{A}$ and $SR_{B}$ are tuned equal, the complete
apparatus forms a balanced MZ interferometer, so that $\psi _{i}^{0}$ ($\psi
_{i}^{1}$) will make the detector $D_{0}$ ($D_{1}$) click with certainty in
principle.


\begin{figure*}[tbp]
\centering
\includegraphics[scale=0.9]{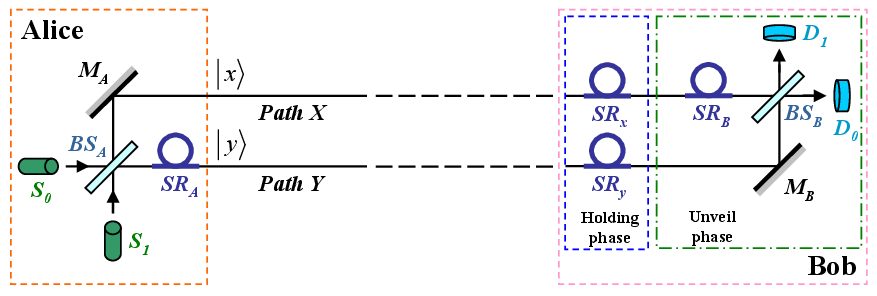}
\caption{Diagram of the experimental apparatus of our QBC protocol. Alice
sends photons from the single-photon source $S_0$ ($S_1$) when she wants to
commit $b=0$ ($b=1$). Both $BS_{A}$\ and $BS_{B}$\ are
\textcolor{black}{50:50 non-polarizing beam splitters}, and $M_{A}$, $M_{B}$\ are mirrors. $SR_{A}$%
, $SR_{B}$, $SR_{x}$ and $SR_{y}$ are storage rings. The photons are
finally detected by the detectors $D_{0}$ and $D_{1}$.}
\label{fig:epsart}
\end{figure*}

Before running the protocol, Bob should setup another set of devices at his
own site as a reference, which is completely identical to that of Alice's.
By sending photons using this reference set and monitoring his detectors $%
D_{0}$ and $D_{1}$, he can estimate the error rate $\varepsilon $ of the
whole system, i.e., the probability that the photon $\psi _{i}^{0}$ ($\psi
_{i}^{1}$) sent from the source $S_{0}$ ($S_{1}$) will mistakenly make the
detector $D_{1}$ ($D_{0}$) click or simply get lost. For better performance,
if the distance between Alice and Bob is very long, paths $X$ and $Y$ in
Fig. 1 should be implemented using optical fibers, instead of letting the
photons travel through free space. Meanwhile, Bob should also have optical
fibers of the same length in his reference set, and place them in an
environment (e.g., temperature, humidity, etc.) similar to that of the optical
fibers placed between Alice and Bob in the actual set. The purpose is to
ensure that the error rate $\varepsilon $ that Bob
\textcolor{black}{learns} from his
reference set is very close to the one in the actual set. After obtaining $%
\varepsilon $ all by himself, Bob runs the following experimental protocol
with Alice.

\bigskip

\textit{Our experimental QBC protocol:}

\textit{The commit phase:}

\textit{(i) Alice and Bob agree on a maximum delay time }$\tau _{\max }$%
\textit{\ and the sending times }$t_{j}$\textit{\ (}$j=1,...,s$\textit{)
with }$t_{1}<t_{2}<...<t_{s}$\textit{\ and }$\tau _{\max }<t_{j}-t_{j-1}$%
\textit{\ (}$j=2,...,s$\textit{). Then Alice decides on the value of }$b$%
\textit{\ (}$b=0$\textit{\ or }$1$\textit{)\ that she wants to commit, and
for }$j=1$\textit{\ to }$s$\textit{:}

\textit{Alice randomly picks }$\tau _{j}\in \lbrack 0,\tau _{\max }]$\textit{%
, and sets the delay time of her storing ring }$SR_{A}$\textit{\ as }$%
\tau _{j}$\textit{. Then she sends Bob a photon }$\Psi _{j}$\textit{\ from
the source }$S_{b}$\textit{\ at time }$t_{j}$\textit{.}

\textit{Note that in each round, }$\tau _{j}$\textit{\ is independently
chosen, while }$b$\textit{\ remains the same for all }$j$\textit{.}

\textit{The holding phase:}

\textit{(ii) Bob stores the wave packets of each photon in }$SR_{x}$\textit{%
\ and }$SR_{y}$\textit{\ unmeasured.}

\textit{The unveil phase:}

\textit{(iii) Alice announces the values of }$b$\textit{\ and all }$\tau
_{j} $\textit{\ (}$j=1,...,s$\textit{).}

\textit{(iv) For }$j=1$\textit{\ to }$s$: \textit{Bob sets the delay time of
his storing ring }$SR_{B}$\textit{\ as }$\tau _{j}$\textit{. Then he
releases the wave packets of photon }$\Psi _{j}$\textit{\ from }$SR_{x}$%
\textit{\ and }$SR_{y}$\textit{\ and directs them into his part of the MZ
interferometer (as presented in the green dash-dot box at the right-hand
side of Fig. 1).}

\textit{If there are totally about }$(1-\varepsilon )s$\textit{\ photons
(see Appendix A for the tolerable range of statistical deviations) detected by
}$D_{b}$\textit{\ instead of }$D_{\bar{b}}$\textit{ then Bob accepts
Alice's commitment. Otherwise Bob concludes that Alice cheated.}

\bigskip

\subsection{The relationship between the two protocols}

Now we show that in principle, the above experimental protocol is a faithful
implementation of the theoretical one. In the experimental protocol,
following the occupation number representation widely used in quantum optics
\cite{qi858}, at time $t$\ if there is a wave packet of a photon on path $%
X $ and no wave packet on path $Y$, the state can be denoted by $\left\vert
1\right\rangle _{X}\left\vert 0\right\rangle _{Y}$. Or if there is a wave
packet on path $Y$ and no wave packet on path $X$, the state can be denoted
by $\left\vert 0\right\rangle _{X}\left\vert 1\right\rangle _{Y}$. To make
the time $t$\ more explicit, let us write them as $\left\vert t\right\rangle
_{X}\left\vert 0\right\rangle _{Y}$\ and $\left\vert 0\right\rangle
_{X}\left\vert t\right\rangle _{Y}$, respectively. That is, we use the
symbol $t$\ in $\left\vert ...\right\rangle $ to denote the time that the
wave packet of a single photon presents in the path, instead of the number
of photons; and $\left\vert 0\right\rangle $\ means that no wave packet is
presented in the path at any time. Obviously, the state $\left\vert
t\right\rangle _{P}$\ is orthogonal to $\left\vert t^{\prime }\right\rangle
_{P}$\ ($P=X,Y$) for any $t\neq t^{\prime }$ and they are all orthogonal to $%
\left\vert 0\right\rangle _{P}$. For simplicity, suppose that except for $%
SR_{A}$,\ $SR_{B}$,\ $SR_{x}$,\ and $SR_{y}$, the time for the photon to
travel through all other devices in Fig. 1 is negligible. Under this
formalism, when Alice sends the photon $\Psi _{j}$ ($j=1,...,s$) from the
source $S_{b}$\ at time $t_{j}$, the initial state of $\Psi _{j}$ after
passing $BS_{A}$\ is%
\begin{equation}
\left\vert \Psi _{j}\right\rangle _{ini}=\frac{1}{\sqrt{2}}(\left\vert
t_{j}\right\rangle _{X}\left\vert 0\right\rangle _{Y}+(-1)^{b}\left\vert
0\right\rangle _{X}\left\vert t_{j}\right\rangle _{Y}).  \label{initial}
\end{equation}%
After passing $SR_{A}$ which introduces the delay time $\tau _{j}$\ to path $%
Y$, the state of $\Psi _{j}$ that
\textcolor{black}{left} Alice's site is%
\begin{equation}
\left\vert \psi _{j}\right\rangle _{A}=\frac{1}{\sqrt{2}}(\left\vert
t_{j}\right\rangle _{X}\left\vert 0\right\rangle _{Y}+(-1)^{b}\left\vert
0\right\rangle _{X}\left\vert t_{j}+\tau _{j}\right\rangle _{Y}).
\label{send}
\end{equation}

In the unveil phase when Bob learns Alice's delay time $\tau _{j}$\ and sets
$SR_{B}$ accordingly, the final state of the photon $\Psi _{j}$ arriving at $%
BS_{B}$\ after passing $SR_{x}$,\ $SR_{y}$ and $SR_{B}$ is%
\begin{eqnarray}
\left\vert \psi _{j}\right\rangle _{fin} &=&\frac{1}{\sqrt{2}}(\left\vert
t_{j}+{\large \tau }_{hold}+\tau _{j}\right\rangle _{X}\left\vert
0\right\rangle _{Y}  \nonumber \\
&&+(-1)^{b}\left\vert 0\right\rangle _{X}\left\vert t_{j}+\tau _{j}+{\large %
\tau }_{hold}\right\rangle _{Y})  \nonumber \\
&=&\frac{1}{\sqrt{2}}(\left\vert t_{j}^{\prime }\right\rangle _{X}\left\vert
0\right\rangle _{Y}+(-1)^{b}\left\vert 0\right\rangle _{X}\left\vert
t_{j}^{\prime }\right\rangle _{Y}),  \label{final}
\end{eqnarray}%
where ${\large \tau }_{hold}$\ is the length of the time that $\Psi _{j}$
was stored in $SR_{x}$ and\ $SR_{y}$, and%
\begin{equation}
t_{j}^{\prime }\equiv t_{j}+{\large \tau }_{hold}+\tau _{j}.
\end{equation}

Meanwhile, when combined with $BS_{B}$, the detectors $D_{0}$ and $D_{1}$
serve as the projective operators%
\begin{equation}
P_{0}\equiv \left\vert \psi \right\rangle _{0}\left\langle \psi \right\vert
_{0}  \label{p0}
\end{equation}%
and%
\begin{equation}
P_{1}\equiv \left\vert \psi \right\rangle _{1}\left\langle \psi \right\vert
_{1},  \label{p1}
\end{equation}%
respectively, where%
\begin{equation}
\left\vert \psi \right\rangle _{0}\equiv \frac{1}{\sqrt{2}}(\left\vert
t_{B}\right\rangle _{X}\left\vert 0\right\rangle _{Y}+\left\vert
0\right\rangle _{X}\left\vert t_{B}\right\rangle _{Y})
\end{equation}%
and%
\begin{equation}
\left\vert \psi \right\rangle _{1}\equiv \frac{1}{\sqrt{2}}(\left\vert
t_{B}\right\rangle _{X}\left\vert 0\right\rangle _{Y}-\left\vert
0\right\rangle _{X}\left\vert t_{B}\right\rangle _{Y})
\end{equation}%
with $t_{B}$\ denoting the time that Bob applies the measurement. Therefore,
if Bob takes $t_{B}=t_{j}^{\prime }$, then in the ideal case where the error
rate $\varepsilon $ is negligible, the detector $D_{b}$ should click with
certainty where $b$ is Alice's committed bit. Otherwise he knows that Alice
\textcolor{black}{cheated}.

To see that the above presentation of the states is equivalent to that in
our theoretical QBC protocol, let us view the time range $[0,\tau _{\max }]$%
\ (within which Alice picks her delay time $\tau _{j}$) as a series of time
\textcolor{black}{slots} $T_{1}$, $T_{2}$, ..., $T_{i}$, ..., $T_{n-1}$. Here $0\leq T_{i}\leq
\tau _{\max }$ ($i=1,...,n-1$), and $T_{i}\neq T_{i^{\prime }}$ for any $%
i\neq i^{\prime }$. When time can be treated as a continuous variable, there
is an infinite number of choices for $T_{i}$, i.e., $n\rightarrow \infty $.
Now for each $\Psi _{j}$ ($j=1,...,s$),\ let us define%
\begin{equation}
\left\vert 0\right\rangle \equiv \left\vert t_{j}\right\rangle
_{X}\left\vert 0\right\rangle _{Y}  \label{0}
\end{equation}%
and%
\begin{equation}
\left\vert i\right\rangle \equiv \left\vert 0\right\rangle _{X}\left\vert
t_{j}+T_{i}\right\rangle _{Y}.  \label{i}
\end{equation}%
It is easy to verify that $\left\langle i^{\prime }\right. \left\vert
i\right\rangle =\delta _{i^{\prime }i}$. That is, a single photon $\Psi _{j}$
can be treated as an $n$-dimensional system, with $\{\left\vert
i\right\rangle ,i=0,...,n-1\}$ being an orthonormal basis.

With these newly defined $\left\vert 0\right\rangle $\ and $\left\vert
i\right\rangle $, we can see that in the experimental protocol, when Alice
chooses the delay time as $\tau _{j}=T_{i}$ ($i\in \{1,...,n-1\}$), Eq. (\ref%
{send}) can be rewritten as%
\begin{equation}
\left\vert \psi _{j}\right\rangle _{A}=\frac{1}{\sqrt{2}}(\left\vert
0\right\rangle +(-1)^{b}\left\vert i\right\rangle ).
\end{equation}%
This is exactly the state that Alice sends in step (i) of the theoretical
protocol for committing the bit $b$, as shown by Eqs. (\ref{state+}) and (%
\ref{state-}). Thus, it is proven that our proposed experimental protocol is
equivalent to the theoretical one in principle, so that the security proof
in
\textcolor{black}{Ref.} \cite{HePRSA} also applies. Consequently, the experimental protocol is
secure as long as time can be treated as a continuous variable so that the
condition $n\rightarrow \infty $ can be reached.

\subsection{Feasibility}

The \textcolor{black}{experimental} apparatus shown in Fig. 1 is much the same as those of the
quantum key distribution (QKD) and quantum private query protocols in Refs. \cite%
{qi858,qi917,qi1573}. The only difference is that our protocol requires two
more storage rings $SR_{x}$ and\ $SR_{y}$. The QKD protocol in Ref. \cite{qi858}
was already realized experimentally in Ref. \cite{qi889}. By comparing our
apparatus with Fig. 1 of Ref. \cite{qi889}, we can see that the technology in
Ref. \cite{qi889} is sufficient for implementing our protocol too. Detailed
description of the actual experimental devices can be found in Section III
of Ref. \cite{qi889}.

An important part \textcolor{black}{of the implementation} is to find storage rings $SR_{x}$\ and $SR_{y}$\
with a sufficiently long delay time, because they determine the holding time
(the time interval between the commit phase and the unveil phase) of the
protocol. Using $150km$ optical fiber (which was proven to be able to
guarantee sufficiently high key rate for QKD in practice) to make the
storage ring can generate about $500\mu s$\ delay time. While such a holding
time seems short, it is already sufficient for practical applications such
as quantum coin tossing, as shown in Appendix B.

\subsection{Practical difficulties}

The security of the practical implementation of the protocol, however, is
limited by two difficulties. (1) The length of the classical communications
between Alice and Bob has to remain finite, so that Alice cannot announce
the delay times $\tau _{j}$\ ($j=1,...,s$)
\textcolor{black}{with} an unlimited number of digits.
(2) The delay time of the storage rings $SR_{A}$ and\ $SR_{B}$ cannot be
adjusted to an unlimited precision either, so that Alice and Bob cannot set $%
\tau _{j}$\ precisely to any desired value. Consequently, the number of
choices for the time slots $T_{i}$ in Eq. (\ref{i}) (from which $\tau _{j}$\
can be picked) cannot really go to infinite. Instead, when the above two
difficulties limit the precision of time control to $\Delta \tau $, the
number of time slots within the range $[0,\tau _{\max }]$\ is%
\begin{equation}
n=\tau _{\max }/\Delta \tau +1.
\end{equation}%
Therefore, the quantum optical method in Section II B actually implements
finite $n$-dimensional systems, instead of infinite-dimensional ones.
That is, though the two protocols are equivalent in principle, in practice the experimental scheme is not a faithful implementation of our
theoretical protocol in Section II A. Thus, it cannot be as secure as the
latter.

Nevertheless, making use of this limitation for cheating is technically
challenging too. Suppose that Alice and Bob know the value of $\Delta \tau $
and therefore know the actual finite $n$. According to Section 5 of Ref.
\cite{HePRSA}, if Alice wants to cheat, she needs to have the technology to
prepare entangled states in the form of Eq. (\ref{omega}) in the commit
phase, which is the quantum superposition of $n$ different states. Moreover,
later in the unveil phase when she measures system $\alpha $ to complete her
cheating, she needs to discriminate her measurement result between $%
\left\vert \alpha _{i+}\right\rangle $ and $\left\vert \tilde{\alpha}%
_{i-}\right\rangle $, where the latter is defined by \textcolor{black}{Eq. (\ref{alfa-})}.
multiplying $\left\langle \alpha _{i+}\right\vert $\ by it and we yield%
\begin{equation}
|\left\langle \alpha _{i+}\right. \left\vert \tilde{\alpha}%
_{i-}\right\rangle |^{2}=1-\frac{4}{n+2}  \label{old eq31}
\end{equation}%
(i.e., Eq. (3.10) of Ref. \cite{HePRSA}). In our experimental protocol,
suppose that the storage rings can achieve a precision of $\Delta \tau
=300ps $; then for $\tau _{\max }=500\mu s$, there is $n\simeq 1.67\times
10^{6}$. We can see that $|\left\langle \alpha _{i+}\right. \left\vert
\tilde{\alpha}_{i-}\right\rangle |^{2}$ is so close to $1$, that even a tiny
bit of noise and error in Alice's experimental devices (which is inevitable
in practice) could make the discrimination between $\left\vert \alpha
_{i+}\right\rangle $ and $\left\vert \tilde{\alpha}_{i-}\right\rangle $
impossible. On the other hand, if Bob is dishonest and wants to learn the
committed bit $b$ before the unveil phase, according to Section 7 of Ref.
\cite{HePRSA} he needs to be capable of discriminating the two density
matrices $\rho _{0}^{\beta }=\rho _{+}^{\otimes s}$ and $\rho _{1}^{\beta
}=\rho _{-}^{\otimes s}$ where the trace distance between $\rho _{+}$ and $%
\rho _{-}$ is%
\begin{equation}
D(\rho _{+},\rho _{-})=\frac{1}{\sqrt{n-1}}  \label{trace distance}
\end{equation}%
(i.e., Eq. (3.1) of Ref. \cite{HePRSA}). Again, in the practical setting,
such a tiny difference between the states could be completely drown by the
noise and error in the experimental devices. Therefore, even with the
limited $n$ value achievable today, the experimental protocol in Section II B
can at least be used as a practically secure (instead of unconditionally
secure) QBC scheme, or serve as a proof-of-principle implementation of the
theoretical protocol in Section II A.

\section{Discussion}

In summary, we showed that as long as time can be treated as a continuous variable, then
each infinite-dimensional system in the unconditionally secure QBC protocol
proposed in Ref. \cite{HePRSA} can be realized using a single photon. Thus
we obtained an experimental implementation of this QBC protocol which is
feasible under currently available technology.
\textcolor{black}{Other \textquotedblleft
post-cold-war era\textquotedblright\ multi-party cryptographic protocols are therefore made possible too, e.g., quantum coin tossing, as elaborated in Appendix B.}

The dimension of the systems cannot really be infinite in practice though,
making \textcolor{black}{the current} experimental implementation a practically secure QBC only. But
it still has an advantage over many other practically secure QBC protocols
(e.g., Refs. \cite{qbc59,qbc65,qbc219}). While these protocols could be more
feasible than ours in practice, their security is based on certain practical
limitations. For example, currently available quantum memory cannot store
the quantum states for a long period of time, so that as long as the holding
phase of the protocol is longer than this period of time, we can be sure
that Alice can no longer cheat by storing the states and delay the measurement until the
unveil phase. But as technology advances, the storage time
\textcolor{black}{of} quantum
memory will increase, making it harder and harder to keep the
\textcolor{black}{corresponding} protocol secure. On the contrary, the security of our experimental protocol
is based on the unconditionally secure theoretical protocol in Ref. \cite%
{HePRSA}. Practical limitation is the reason that weakens its security so
that it can be practically secure only, not the reason that makes it secure.
Therefore, with the advance of the technology on the precision of the delay
time adjustment, we can expect the security of this experimental protocol to
be constantly improved towards that of the theoretical protocol in Ref. \cite%
{HePRSA}.
\textcolor{black}{Meanwhile, it is also worth studying whether some new technologies can be adopted to implement the infinite-dimensional systems to make our protocol even more feasible, e.g., the continuous phase noise resulting from gain switching laser operation \cite{phase}.}

\color{black}

Our result may also contributes to the development of fundamental
theories. There is a brilliant idea called the CBH theorem \cite{qi256}%
, which intents to deduce quantum theory by using three information-theoretic
constraints as fundamental axioms. (I) the impossibility of superluminal information transfer, (II) the impossibility
of perfectly broadcasting of an unknown state, and (III) the impossibility
of unconditionally secure BC. The reason for including the last constraint, is that Alice's cheating strategy against BC requires the use of entangled states, as can be seen from Section II A of the current work. That is, the impossibility
of unconditionally secure BC entails the existence of non-locality, which is one of the essential feature of quantum theory. However, in the three QBC protocols we previously proposed in Refs. \cite{HePRA,HeJPA,HeQIP} which manage to evade the MLC no-go theorem, non-locality is necessary even for honest participants. This observation implies that if the constraint (III) is wrong, i.e., unconditionally secure QBC exists, then non-locality is also entailed. For this reason, we tend to believe that the (in)validity of the constraint (III) has nothing to do with the existence of non-locality. The latter has to exist in our physical world anyway. To complete the deduction of quantum theory from information-theoretic axioms, we should look for another constraint to replace constraint (III). Nevertheless, the finding of the unconditionally secure QBC protocol in Ref. \cite{HePRSA} may blur the above picture at first glance. This is because the protocol makes use of infinite-dimensional systems instead of entangled states, so that it seems to indicate that besides non-locality, infinity should also be taken into account as the quantum resources that need to be entailed if we want to build quantum theory completely on top of information-theoretic axioms.
But the result in the current work provides a clue to clean this mist. As can be seen from Fig. 1, in our implementation of the infinite-dimensional systems, each photon state is divided spatially into two wave packets that travel along different paths, so that non-locality is introduced. Therefore, the current implementation scheme bridges infinity with non-locality, so that non-locality could still be considered as the only quantum resource that guarantees unconditionally secure QBC.

\color{black}



\vspace{6pt}


\noindent
\textbf{Funding:}
This work was supported in part by Guangdong Basic and Applied Basic
Research Foundation under grant No. 2019A1515011048.



\appendix
\section[\appendixname~\thesection]{Confidence interval of the error rate}
In the last step of our experimental QBC protocol, Bob is supposed to find $%
(1-\varepsilon )s$\ photons detected by $D_{b}$\ instead of $D_{\bar{b}}$
when Alice is honest. But since $\varepsilon $\ is only the statistical
average of the error rate of the experimental apparatus, Bob cannot expect
to find exactly $(1-\varepsilon )s$\ photons detected correctly. A certain
range of statistical deviations has to be allowed. Now let us estimate the
size of this range.

According to Theorem 3.3 of Ref. \cite{qi43} (which is based on Bernshtein's law
of large numbers), when each of the $s$ photons stands the probability $%
p=(1-\varepsilon )$ to make the correct detector $D_{b}$\ click, for
arbitrarily small positive value $\delta \leq p(1-p)$, the probability for
the case $\left\vert s^{\prime }/s-p\right\vert \geq \delta $\ to occur
satisfies%
\begin{equation}
\Pr (\left\vert \frac{s^{\prime }}{s}-p\right\vert \geq \delta )\leq
2e^{-s\delta ^{2}}.
\end{equation}%
Here $s^{\prime }$ denotes the actual number of\ photons
\textcolor{black}{being} detected correctly
in a complete run of the protocol. This inequality means that $s^{\prime }$\
should be within the range $[(1-\varepsilon )s-\delta s,(1-\varepsilon
)s+\delta s]$, except with a probability $2e^{-s\delta ^{2}}$ at the most.
For example, when $s=10000$, $\varepsilon =10\%$ and $\delta =5\%$, we have $%
2e^{-s\delta ^{2}}\simeq 2.8\times 10^{-11}$, which is extremely small. As a
result, the number of photons that
\textcolor{black}{are} actually detected by Bob's $D_{b}$\
should be within the range $[8500,9500]$, otherwise he can confidently
conclude that Alice is dishonest.

\section[\appendixname~\thesection]{Quantum coin tossing as an application}
Although the holding time of our QBC protocol may look short
\textcolor{black}{even} with
state-of-the-art optical delay devices, it is sufficient for some practical
applications. Here, as an example, we will show how quantum coin tossing
(QCT, a.k.a. quantum coin flipping) \cite{qi365} can be realized.

The goal of QCT is to provide a method for two separated parties Alice and
Bob to generate a random bit value $c=0$ or $1$ remotely, while they do not
trust each other. If the parties have opposite desired values, e.g., Alice
wants $c=0$ while Bob wants $c=1$, then it is called weak QCT. Or if their
desired values are random, then it is called strong QCT. Here we focus on
strong QCT. Such a protocol is considered secure if neither party can bias
the outcome, so that $c=0$ and $c=1$ will occur with the equal
probabilities $1/2$, just as if they are tossing an ideal fair coin.
Possible application scenario of QCT may include the case where divorced and
separated couples, who want to decide how to divide their property through
telephone. Other more complicated applications such as online gambling can
be constructed through it too.

Strong QCT with an arbitrarily small bias was also considered a hard task if
unconditionally secure QBC is impossible \cite{qi58}. But when QBC becomes
available, QCT can easily be built upon it as follows.

\bigskip

\textit{Strong QCT protocol:}

\textit{(I) Alice and Bob complete the commit phase of our QBC protocol,
where Alice picks the value of her committed bit }$b$\textit{\ randomly.}

\textit{(II) During the holding phase, Bob picks a random bit }$x$\textit{\
and announces it to Alice through the classical channel.}

\textit{(III) Alice and Bob complete the unveil phase of our QBC protocol.
That is, Alice unveils her committed bit }$b$\textit{, and Bob checks
whether she is honest or not.}

\textit{(IV) Both Alice and Bob accept }$y\equiv b\oplus x$\textit{\ as the
coin tossing result.}

\bigskip

It is trivial to show that if the QBC protocol is secure (i.e., Alice cannot
change $b$ after the commit phase, and Bob cannot know $b$ before the unveil
phase), then the value of the final tossing result $y$ is completely random.
Neither Alice nor Bob can bias it to any specific value.

In this example, the commit and unveil phases of the QBC protocol are
separated merely by step (II), where only one classical bit $x$ is
transferred. Bob can decide the value of $x$ beforehand but keep it secret
from Alice during step (I). Then step (II) can be performed automatically
under the control of
\textcolor{black}{classical} computers, which can be done very fast. Therefore, the
holding time in our QBC protocol is already long enough for such operations,
so that it can result in a useful QCT protocol in practice.



\end{document}